\documentclass[amssymb,aps,twocolumn,showpacs,nofootinbib]{revtex4}
\usepackage[]{graphicx}
\def\ket#1{| #1 \rangle}
\def\bra#1{\langle #1 |}
\def\bracket#1#2{\langle #1 | #2 \rangle}

\begin{document}

\title{Unconditional Security of a Three State Quantum Key Distribution Protocol}

\author{J.-C. Boileau$^1$, K. Tamaki$^2$, J. Batuwantudawe$^1$, R. Laflamme$^{1,2}$, J. M. Renes$^{3,4}$}

\affiliation{$^1$Institute for Quantum Computing, University of Waterloo, Waterloo, ON, N2L 3G1, Canada.\\
$^2$Perimeter Institute for Theoretical Physics, 35 King Street North, Waterloo, ON, N2J 2W9, Canada.\\
$^3$Department of Physics and Astronomy, University of New Mexico, Albuquerque, NM 87131--1156, USA\\
$^4$IAKS, Arbeitsgruppe Quantum Computing, Universit\"at Karlsruhe, Am Fasanengarten 5, D-76131 Karlsruhe, Germany}

\date{\today}

\begin{abstract}

Quantum key distribution (QKD) protocols are cryptographic techniques with security based only on the laws of quantum mechanics. Two prominent QKD schemes are the BB84 and B92 protocols that use four and two quantum states, respectively.  In 2000, Phoenix {\it et al.} proposed a new family of three state protocols that offers advantages over the previous schemes. Until now, an error rate threshold for security of the symmetric trine spherical code QKD protocol has only been shown for the trivial intercept/resend eavesdropping strategy. In this paper, we prove the unconditional security of the trine spherical code QKD protocol, demonstrating its security up to a bit error rate of 9.81\%. We also discuss on how this proof applies to a version of the trine spherical code QKD protocol where the error rate is evaluated from the number of inconclusive events.
 
\end{abstract}


\maketitle

Quantum key distribution (QKD) protocols permit two separated parties, say Alice and Bob, to construct a secret shared string of bits that may be used for cryptography. The first QKD protocol, called BB84, was invented by Bennett and Brassard in 1984~\cite{BB84}. It requires Alice to randomly produce four different states $\ket0$, $\ket1$, $\ket+$, $\ket-$ and send them through a quantum channel to Bob who measures them randomly in the $\{\ket0, \ket1\}$ basis or in its conjugate basis $\{\ket+, \ket-\}$. The unconditional security of this protocol was first shown by Mayers in 1996~\cite{M96}. A simpler QKD protocol, B92, was proposed by Bennett in 1992~\cite{B92}. It requires Alice to produce only two non-orthogonal states, say $\ket{\psi_1}$ and $\ket{\psi_2}$, and Bob to perform the measurement described by the POVM  (positive operator-valued measure) $\{\alpha\ket{\overline{\psi}_1}\bra{\overline{\psi}_1}, \alpha\ket{\overline{\psi}_2}\bra{\overline{\psi}_2}, \openone - \alpha\ket{\overline{\psi}_1}\bra{\overline{\psi}_1}-\alpha\ket{\overline{\psi}_2}\bra{\overline{\psi}_2} \}$, where $\ket{\overline{\psi}_1}$ and $\ket{\overline{\psi}_2}$ are orthogonal to $\ket{\psi_1}$ and $\ket{\psi_2}$, respectively, and $\alpha$ equals $\frac{1}{1+|\bracket{\psi_1}{\psi_2}|}$ to optimize the probability of a conclusive result to occur. Recent results by Tamaki \emph{et al.} showed that B92 is secure for small noise, and the security threshold depends on qubit losses~\cite{TKI03TL04}.

Phoenix \emph{et al.}~\cite{PBC00} postulated that the addition of a third state to the B92 protocol could considerably enhance its security and would be optimal if the three quantum states form an equilateral triangle on the X-Z plane in the Bloch sphere. We call this particular case the trine spherical code QKD protocol or the PBC00 protocol. PBC00 is similar to B92, except that Alice randomly chooses two of three states for a basis instead of using two fixed states. From Eve's point of view, the state sent by Alice is a maximally mixed state, unlike in B92. This feature is similar to BB84, in which the choice of encoding basis by Alice corresponds to a random rotation  (the identity or the Hadamard transformation).  In PBC00, the choice of encoding basis by Alice also corresponds to a rotation---by 120 degrees, 240 degrees, or none at all. Intuitively, we could expect to find a security threshold for PBC00 that is independent of qubit losses and that is close to the one of BB84. As we will explain in detail, our security proof also applies to a slightly modified version of the PBC00 protocol proposed by Renes~\cite{R04v2} that we will refer to as R04. In this protocol, the error rate is estimated from the number of inconclusive events, and all conclusive results can be used as data bits instead of wasting some as test bits. This also simplifies the classical communication between Alice and Bob because they do not need to randomly select a set of test bits and broadcast them.

Up to now, the high error rate threshold for the security of the PBC00 protocol has only been shown in the special case of the intercept/resend attack~\cite{R04v2}. In this Letter, we will give a proof of the unconditional security of the PBC00 and R04 protocols. Assuming one-way classical communication, we show that these protocols are secure up to a bit error rate of 9.81\%. In order to establish security, we first propose a QKD scheme based on an Entanglement Distillation Protocol (EDP)~\cite{EDP}. This protocol uses a Calderbank, Shor, and Steane (CSS) code~\cite{CSS}, a technique first used by Shor and Preskill in their security proof for BB84~\cite{SP00}. Before running an EDP based on CSS codes, Alice and Bob perform state rotations followed by Bob's local filtering operation (LF)~\cite{Filtering}. The local filtering operation correlates the phase and bit error rates, as in the security proof of B92~\cite{TKI03TL04}. Thanks to the state rotation by Alice and Bob, we achieve phase error estimation from bit error estimation. We will also explain how the security of R04 follows from that of PBC00.  

The PBC00 protocol involves three states $\ket{\psi_1}\equiv\frac{1}{2}\ket{0_x}+\frac{\sqrt3}{2}\ket{1_x}$, $\ket{\psi_2}\equiv\frac{1}{2}\ket{0_x}-\frac{\sqrt3}{2}\ket{1_x}$, and $\ket{\psi_3}\equiv\ket{0_x}$, where $\{\ket{0_x},\ket{1_x}\}$ is a basis state ($X$-basis) of a qubit state. The $Z$-basis is defined by $\{\ket{j_z}\equiv[\ket{0_x}+(-1)^j\ket{1_x}]/\sqrt{2}\}$ ($j=0,1$), and we also define states $\ket{\bar{\psi}_1} = \frac{\sqrt3}{2}\ket{0_x}-\frac{1}{2}\ket{1_x}$, $\ket{\bar{\psi}_2} =  \frac{\sqrt3}{2}\ket{0_x}+\frac{1}{2}\ket{1_x}$ and $\ket{\bar{\psi}_3} = \ket{1_x}$ that are orthogonal to $\ket{\psi_1}$, $\ket{\psi_2}$, and $\ket{\psi_3}$, respectively. The protocol proceeds as follows.
\bigskip

\noindent{\bf PBC00:}\\
\noindent{\bf 1.1} Alice creates a large trit string $r$ and a large bit string $b$ of the same length.  For each $r_i$, the $i^{th}$ trit value of the trit string $r$, she chooses the set $\{\ket{\psi_1}, \ket{\psi_2}\}$ (if $r_i=0$), $\{\ket{\psi_2}, \ket{\psi_3}\}$ (if $r_i=1$), and $\{\ket{\psi_3}, \ket{\psi_1}\}$ (if $r_i=2$). If the $i^{th}$ bit value $b_i$ is 0, she prepares the first state of the chosen pair. If the bit is 1, she prepares the second state. Alice sends all prepared qubits to Bob.

\noindent{\bf 1.2} 
Bob performs a measurement described by the POVM $\{\frac{2}{3}\ket{\bar{\psi}_1}\bra{\bar{\psi}_1}, \frac{2}{3}\ket{\bar{\psi}_2}\bra{\bar{\psi}_2}, \frac{2}{3}\ket{\bar{\psi}_3}\bra{\bar{\psi}_3}
\}$. 
He publicly announces when all his measurements are done, and Alice in turn announces the trit string $r$.

\noindent{\bf 1.3} Bob regards the $i^{th}$ measurement outcome $\ket{\bar{\psi}_1}$ (if $r_i=0$), $\ket{\bar{\psi}_2}$ (if $r_i=1$), and $\ket{\bar{\psi}_3}$ (if $r_i=2$) as the bit value 0. Similarly, he regards $\ket{\bar{\psi}_2}$ (if $r_i=0$), $\ket{\bar{\psi}_3}$ (if $r_i=1$), and $\ket{\bar{\psi}_1}$ (if $r_i=2$) as the bit value 1. All other events are regarded as inconclusive. Bob announces whether his measurement outcome is inconclusive or not. Alice and Bob keep all data where Bob's outcome is conclusive, discarding the rest.

\noindent{\bf 1.4} Alice randomly chooses half of the remaining events as test bits in order to estimate the bit error rate on the code bits, and announces her selection to Bob. They compare the values of their test bits, aborting the protocol if the error rate is too high.

\noindent{\bf 1.5} By public discussion, they run classical error correction and privacy amplification protocols to share a secure secret key.
\bigskip

In order to prove the security of PBC00, we relate this protocol to a secure QKD based on an EDP initiated by state rotations and a LF, followed by error correction using CSS codes~\cite{SP00}. The LF is designed so that it probabilistically distills the maximally entangled state $\ket{\Phi^{+} }= \frac{1}{\sqrt2}(\ket{0_z}\ket{0_z}+\ket{1_z}\ket{1_z})$ if the filtering succeeds. Thus, the successful local filtering operation can be written by a Kraus operator $F=\ket{0_x}\bra{0_x}+\frac{1}{\sqrt3}\ket{1_x}\bra{1_x}$. For later convenience, we define $\ket{\Phi^{-}}=\frac{1}{\sqrt2}(\ket{0_z}\ket{0_z}-\ket{1_z}\ket{1_z})$, $\ket{\Psi^\pm}=\frac{1}{\sqrt2}(\ket{0_z}\ket{1_z}\pm\ket{1_z}\ket{0_z})$, and $R_y(2b\pi/3)$ as a $2b\pi/3$ rotation around the Y-axis in the Bloch sphere. The following is the secure QKD based on an entanglement distillation protocol that will be reduced to PBC00.
\bigskip

\noindent{\bf QKD based on EDP:}

\noindent{\bf 2.1} Alice creates many pairs of qubits in the state $\ket{\phi}=\frac{1}{\sqrt2}(\ket{0_z}_{\rm{A}}\ket{\psi_1}_{\rm{B}}+\ket{1_z}_{\rm{A}}\ket{\psi_2}_{\rm{B}})$, and randomly chooses a large trit string $r$ whose length equals the number of prepared qubit pairs. She applies $R_y(2r_{i}\pi/3)$ on the second qubit of every pair and sends them to Bob. 
 
\noindent{\bf 2.2} Upon receiving the $i^{th}$ signal state, Bob determines whether the signal is in a qubit state or not which physically corresponds to detecting a photon or not. 
If it is, he declares this publicly to Alice, who in turn declares the trit value $r_{i}$. Bob applies $R_y(-2r_{i}\pi/3)$ on that qubit state followed by the filtering operation. In cases where the filtering operation does not succeed or Bob receives a state that is not a qubit state, he publicly tells Alice to discard her corresponding qubits.

\noindent{\bf 2.3} Alice randomly chooses half of the remaining states as test bits and the other half as code bits, and announces her selection to Bob. For the test bits, Alice and Bob each 
measure their halves in the $Z$-basis. By public discussion, they determine the number of bit errors. If the number of errors in the test bits is too high, they abort the protocol.

\noindent{\bf 2.4} By public discussion, Alice and Bob agree on an appropriate CSS code and run the EDP based on the CSS code to distill nearly perfect Bell 
states from the remaining qubit pairs (code pairs).

\noindent{\bf 2.5} Alice and Bob each measure the Bell pairs in the $Z$-basis to obtain a shared secret key. 
\bigskip

First, we reduce this EDP based protocol into the PBC00 protocol. The reduction can be made 
in the manner of Shor and Preskill~\cite{SP00}. Their reduction technique implies that, in the context of QKD, the EDP based on CSS codes requires Alice to only perform $Z$-basis measurements immediately after she has prepared the state $\ket{\phi}$ and Bob to only perform $Z$-basis measurements immediately after he has performed the filtering operation. The $Z$-basis measurement, together with Alice's rotation, is equivalent to the situation where Alice randomly sends $\ket{\psi_1}$, $\ket{\psi_2}$, or $\ket{\psi_3}$ to Bob. 
On Bob's side, the rotation followed by the filtering operation and $Z$-basis measurement is described by the following POVM,
\begin{eqnarray}
& &R_y(2r_{i}\pi/3)F^{\dagger}\ket{0_z}\bra{0_z}FR_y(-2r_{i}\pi/3),\nonumber\\
& &R_y(2r_{i}\pi/3)F^{\dagger}\ket{1_z}\bra{1_z}FR_y(-2r_{i}\pi/3),\nonumber\\
& &R_y(2r_{i}\pi/3)(\openone-F^{\dagger}F)R_y(-2r_{i}\pi/3),
\end{eqnarray}
which are equivalent as a set to the POVM $\{\frac{2}{3}\ket{\bar{\psi}_1}\bra{\bar{\psi}_1}, \frac{2}{3}\ket{\bar{\psi}_2}\bra{\bar{\psi}_2}, \frac{2}{3}\ket{\bar{\psi}_3}\bra{\bar{\psi}_3}
\}$, 
regardless of the trit value $r_{i}$. Note that failing the filtering operation is 
equivalent to Bob measuring $\ket{\bar{\psi}_j}$ when Alice encoded in the $\{\ket{\psi_{j+1}}, \ket{\psi_{j+2}}\}$ basis in the PBC00 protocol. This completes the reduction.

The equivalence of the two schemes allows us to use the EDP-based protocol to prove the security of PBC00. Security follows by employing a result of Shor and Preskill. They showed that if the estimations of bit and phase error rates on the code pairs are bounded, except for a failure probability that becomes exponentially small as $N$ increases, then Eve's mutual information on the secret key also becomes exponentially small as $N$ increases. Here, $N$ is the number of impure qubit code pairs. Since a large number of test bits yields an exponentially reliable estimation of the bit error rate on the code pairs, we only have to show how to estimate the phase error rate from the bit error rate on the code pairs in our protocol. In the case of BB84, it is trivial to deduce that the phase error rate must equal the bit error rate since the protocol can be interpreted as Alice and Bob measuring all the time in the bit basis, but half the time, the bit errors are inverted with the phase errors. In the case of the three state protocol, the phase error rate is five quarters of the bit error rare. However, the lack of symmetry caused by the filtering operation makes this more difficult to prove.

To make the estimation of the phase error rate
, we appeal to Azuma's inequality~\cite{A67}. 
For a brief explanation of this inequality, consider $N$ random, but dependent events. 
Let  $\{p^{(l)}\}_{l=1,..N}$ be the set of probabilities of having a head in coin flipping for each event. 
Note that $p^{(l)}$ may depend on the results of the $l-1$ previous events. Azuma's inequality tells us that if we perform all the $N$ coin flips and if have $n_{\rm head}$ head events, then the probability that the difference between $n_{\rm head}/N$ and $\frac{1}{N}\sum_{l=1}^{N}p^{(l)}$ is 
larger than some arbitrary small quantity drops exponentially as $N$ increases.

{\bf Proof of our claim:}
 
Definition: {\it Suppose we have a series of events $F_0, F_1, ...$. Let $X_0, X_1, ... $ be random variables. The sequence is a martingale iff the expectation of $X_{i+1}$ conditional to events $F_{i}, F_{i-1}, ... F_{0}$ is equal to $X_i$ for all i.}

Consider the case of $N$ coin tosses, where the probability of getting heads for each coin may be correlated in any way. Consider a series of events $F_0, F_1, ...$. Let $h_i$ be the number of heads from the events $F_{i}, F_{i-1}, ... F_{0}$. Let $X_i$ be $h_i - \sum_{j=1}^{i} p^{(j)}$ where $p^{(j)}$ is the probability of obtaining a head on the $j^{th}$ coin conditional on events $F_{j-1}, F_{j-2}, ... F_{0}$. The expectation of $X_{i+1}$ conditional on events $F_{i}, F_{i-1}, ... F_{0}$ is $h_i-\sum_{j=1}^{i} p^{(j)}$ plus the expectation of obtaining a head on the $i+1$ coin, minus $p^{(i+1)}$. Since the expectation of obtaining a head on the $i+1$ coin minus $p^{(i+1)}$ is zero, the sequence $X_0, X_1, ...$ is a martingale.

Special Case of Azuma's Inequality: {\it Let $X_0, X_1, ...$ be a martingale sequence such that for 
each $k$, $|X_k-X_{k-1}| \le 1$. Then, for all $N \ge 0$ and any $\lambda \ge 0$, 
\begin{eqnarray*}
Pr[|X_N-X_0]\ge \lambda| \le 2 e^{-\frac{\lambda^2}{2N}}.
\end{eqnarray*}
}
In the case of coin flipping introduced above, the condition  $|X_k-X_{k-1}| \le 1$ is obviously satisfied. 
If we let $\lambda = N \epsilon$, then Azuma's inequality implies that 
\begin{eqnarray*}
Pr\left[\left|\frac{h_N- \sum_{j=1}^{N} p^{(j)}}{N}\right|\ge \epsilon\right] \le 2 e^{-\frac{N \epsilon^2 }{2}}\,,
\end{eqnarray*} 
which proves our claim that the probability that the average number of heads differs from $\frac{\sum_{j=1}^{N-1} p^{(j)}}{N}$ by more than an arbitrarily small quantity, $\epsilon$, drops exponentially as $N$ increases.

$\blacksquare$

For the phase error estimations, we define $\{p^{(l)}_{\rm bit}\}_{l=1,..N}$ and $\{p^{(l)}_{\rm phase}\}_{l=1,..N}$ as the sets of probabilities that Alice and Bob 
detect a bit error and a phase error respectively on the $l^{th}$ qubit pair after they have done the same measurements on the $l-1$ previous pairs. Let $e_{\rm bit}$ and $e_{\rm phase}$ be, respectively, the bit and phase error rates that Alice and Bob would have obtained if they had performed bit and phase error measurements on the code pairs. Azuma's inequality tells us that if $Cp_{\rm bit}^{(l)}= p_{\rm phase}^{(l)}$ is satisfied for all $l$ and a particular value of $C$, then we have the exponentially reliable equality $Ce_{\rm bit}= e_{\rm phase}$. Since $e_{\rm bit}$ gets exponentially closer to the bit error 
rate on the test bits, $e_{\rm err}$, we only need to find a value for $C$.

Before we try to obtain $C$, we must assume that Eve can do any coherent attack on all the qubits sent by Alice and that she can use all the ancillary qubits she wants. We will write a general equation for the state of the $l^{th}$ test pair depending on Eve's action. We must be careful to take into account that Alice and Bob's measurement outcomes on the previous $l-1$ test pairs might affect the measurement outcome for $l^{th}$ test pair. Every qubit pair that has passed the filtering operation has undergone Alice's rotation, Eve's global operation and Bob's rotation followed by the filtering operations. The reduced density operator of the $l^{th}$ qubit can be written as $\rho^{(l)}=\frac{1}{3}\sum_{b=0,1,2} 
\ket{\phi^{(l)}_{b}} \bra{\phi^{(l)}_{b}}$, where $\ket{\phi^{(l)}_{b}} = \openone_{A}\otimes \left[FR_y(-2b\pi/3)\hat{E}^{(l)}R_y(2b\pi/3)\right]_{B}\ket{\phi}$, $\ket{\phi}$ is the state created by Alice in step {\bf 2.1} before she applies a rotation, and $\hat{E}^{(l)}$ represents Eve's action restricted to the $l^{th}$ test pair. For simplicity, we will suppose that Eve's action can be written in the form of a single matrix $\hat{E}^{(l)}$ that needs not to be unitary. As it will soon be obvious, our final result still holds in the most general case, where Eve's action on the $l^{th}$ pair is represented by a superoperator satisfying $\sum_i \hat{E}^{(l)\dagger}_i\hat{E}^{(l)}_i \le \openone$. Note that $\hat{E}^{(l)}$ 
may depend on Eve or Alice and Bob's measurement outcome obtained from the previous $l-1$ test pairs. Also note that we summed over the different values of $b$, since $r$ was selected randomly and independently of the other operations done by Alice, Eve, or Bob.

The probability of measuring a bit error on the $l^{th}$ test pair is $p^{(l)}_{\rm bit} = \frac{1}{\zeta^{(l)}}(\bra{\Psi^+}\rho^{(l)}\ket{\Psi^+}+\bra{\Psi^-}\rho^{(l)}\ket{\Psi^-})$ and the probability of measuring a phase error is $p^{(l)}_{\rm phase} = \frac{1}{\zeta^{(l)}}(\bra{\Phi^-}\rho^{(l)}\ket{\Phi^-}+\bra{\Psi^-}\rho^{(l)}\ket{\Psi^-})$, where $\zeta^{(l)}=(\bra{\Phi^+}\rho^{(l)}\ket{\Phi^+}+\bra{\Phi^-}\rho^{(l)}\ket{\Phi^-})+(\bra{\Psi^+}\rho^{(l)}\ket{\Psi^+}+\bra{\Psi^-}\rho^{(l)}\ket{\Psi^-})$ is the probability that the filtering operation succeeds on that qubit. Let us suppose that $c_{11}$, $c_{12}$, $c_{12}$ and $c_{22}$ are the elements of $\hat{E}^{(l)}$ in the X basis where the $c_{ij}$'s are any complex numbers. Then, we easily obtain that $\frac{5}{4}p^{(l)}_{\rm bit}=p^{(l)}_{\rm phase}$. Thus, we have $C=\frac{5}{4}$, and by the previous argument, we conclude that the phase error rate on the code pairs, $e_{\rm phase}$, asymptotically approaches $\frac{5}{4}e_{\rm bit}$. This imply that from the measured bit error rate on the test pairs ($e_{\rm err}$), Alice and Bob can not only get an estimate of the bit error rate on the code pairs ($e_{\rm bit}$), but can also deduce the phase error rate on them ($e_{\rm phase}$). 
If Eve's action is represented by a general superoperator, then the above result still holds by linearity. Note that this argument is valid for any eavesdropping allowed by quantum mechanics because we allow $\{p^{(l)}_{\rm bit}\}_{l=1,..N}$ and $\{p^{(l)}_{\rm bit}\}_{l=1,..N}$ to be arbitrary, including any correlations, and because the $\hat{E}^{(l)}$'s are also arbitrary. Thus, our estimation is applicable to any attack, including coherent attacks.

Since we have the bit and phase error rates, we can calculate the secret key rate. The asymptotically achievable key generation rate for bit error rate $e_{\rm bit}$ and phase error rate $e_{\rm phase}$ is given by $p_{\rm conc}\left[1-h(e_{\rm bit})-h(e_{\rm phase})\right]$, where $h(x)=-x\log_2 x-(1-x)\log_2 (1-x)$ is the binary entropy~\cite{CSS}, and $p_{\rm conc}$ is the probability of conclusive events. In our case, the key generation rate is given by $p_{\rm conc}\left[1-h(e_{\rm err})-h(\frac{5}{4}e_{\rm err})\right]$. From this we find that the PBC00 protocol is secure up to $e_{\rm err}\approx 9.81\%$, for which the key generation rate reaches 0. Contrary to the phase error estimation of B92 over lossy and noisy channel~\cite{TKI03TL04}, this threshold is independent of the qubit losses because, in the previous analysis, we considered only the qubits that survived the filtering operation.

In order to compare the security performance of the three-state protocol with the one of B92, BB84 and the six-state protocol (which is similar the BB84 except that Alice and Bob encode and measure in the X, the Z and also the Y basis), we assume that Eve simulates a depolarizing channel where a qubit state $\rho$ evolves as $(1-p)\rho +\frac{p}{3}\sum_{a=x,y,z}\sigma_{a}\rho\sigma_{a}$. Here, $\sigma_{a}$ is the Pauli operator for $a$ component. It is known that B92, BB84 and the six-state protocol are secure up to $p\approx 3.4\%$ \cite{TKI03TL04}, $p\approx 16.5\%$ \cite{SP00} and $p\approx 19.1\%$ \cite{L01}, respectively, while the three-state protocol is secure up to $p\approx 15.2\%$.

The above security proof also applies to the R04 protocol~\cite{R04v2}. It is similar to the PBC00 protocol, except that the rate of inconclusive events is used to estimate the bit error rate in conclusive events, instead of using test bits in step {\bf 1.5}. In the following, we explain how that is possible
. As a first step, we will make a clear distinction between inconclusive results caused by qubit losses and those caused by qubits that have failed the filtering operation.  From now on, inconclusive events exclude the qubits lost in the channel. 
In PBC00, Alice randomly chooses which basis she uses before sending the state. Without threatening the security, we can modify the protocol so that she sends a random state $\ket{\psi_j}$ and waits until Bob has received it before choosing a basis. For each state, Alice can randomly pick between two bases. The one that she chooses determines which result from Bob's POVM  is inconclusive and which one will induce a ``good" conclusive result (by good, we mean not an error). For Eve, there is no way to differentiate between the one that is inconclusive and the one that induces a ``good" conclusive result. This implies that the number of  ``good" conclusive results approximately equals the number of inconclusive results.  Define $I$ as the fraction of inconclusive results left after discarding the lost qubits. Then, $(1-e_{\rm bit})(1-I)$ is close to $I$, where $e_{\rm bit}$ represent the error rate on all the conclusive events. More precisely, the probability that $e_{\rm bit}$ and $\frac{1-2I}{1-I}$ are different by more than an arbitrary quantity goes exponentially small as the number of received qubits increases. Consequently, Alice and Bob can measure the error rate of the conclusive results by counting the number of inconclusive results. 
Note that the fraction of conclusive results is $p_{\rm conc}=\frac{1}{2-e_{\rm bit}} \geqslant \frac{1}{2}$.

In this Letter, we have proven the unconditional security of the PBC00 and R04 protocols, the latter offering the ability to estimate the bit error rate without sacrificing test bits. Using one-way classical communication, we found an error rate threshold of  9.81\%. As in the case of BB84, two-way classical communication could increase the threshold \cite{GL01}. We believe that Azuma's inequality, used in our security proof, might be useful in other QKD protocol security proofs. Finally, we note that the security proof in this Letter could likely be modified to show the unconditional security of the tetrahedron spherical code recently proposed by Renes \cite{R04v2} or of a new three state QKD protocol robust against collective noise~\cite{BGLPS04}.  
 
The authors wish to thank Daniel Gottesman, Hoi-Kwong Lo and Ashwin Nayak for helpful discussions. 
J.C.B. and R.L. acknowledge support from NSERC and R.L. from ARDA. J.M.R. acknowledges the support of ONR Grant No.~N00014-00-1-0578 and BMBF project No.~01/BB01B.

\end{document}